\documentclass[conference]{IEEEtran}
\IEEEoverridecommandlockouts
\usepackage{cite}
\usepackage{amsmath,amssymb,amsfonts}
\usepackage{algorithmic}
\usepackage{graphicx}
\usepackage{textcomp}
\usepackage{xcolor}
\usepackage{multirow}
\usepackage{subfig} 
\usepackage{booktabs}
\def\BibTeX{{\rm B\kern-.05em{\sc i\kern-.025em b}\kern-.08em
    T\kern-.1667em\lower.7ex\hbox{E}\kern-.125emX}}
\begin{document}

\title{Information-Based Model Discrimination for \\Digital Twin Behavioral Matching}

\author{\IEEEauthorblockN{Jairo Viola and YangQuan Chen}
\IEEEauthorblockA{\textit{MESA Lab}\\
\textit{University of California, Merced}\\
Merced, CA, USA \\
{(jviola,ychen53)}@ucmerced.edu}
\and
\IEEEauthorblockN{Jing Wang}
\IEEEauthorblockA{\textit{College of Automation} \\
\textit{Beijing University of Chemical Technology}\\
Beijing, China \\
jwang@mail.buct.edu.cn}
}

\maketitle

\begin{abstract}
Digital Twin is a breaking technology that allows creating virtual representations of complex physical systems based on updated information of the system and its physical laws. However, making the Digital Twin behavior matching with the real system can be challenging due to the number of unknown parameters in each twin.  Its search can be done using optimization-based techniques, producing a family of models based on different system datasets, so, a discrimination criterion is required to determine the best Digital Twin model. This paper presents an information theory-based discrimination criterion to determine the best Digital Twin model resulting from a behavioral matching process. The information gain of a model is employed as a discrimination criterion. Box-Jenkins models are used to define the family of models for each behavioral matching result. The proposed method is compared with other information-based metrics as well as the $\nu$gap metric. As a study case, the discrimination method is applied to the Digital Twin for a real-time vision feedback infrared temperature uniformity control system. Obtained results show that information-based methodologies are useful for selecting an accurate Digital Twin model representing the system among a family of plants.\\
\end{abstract}


\section{Introduction}
Technologies like Artificial Intelligence (AI), Deep Learning, Data Analytics, or Edge computing can increase the smartness in manufacturing processes and automatic control. One of these technologies is the Digital Twin (DT) that can be defined as a precise, virtual copies of machines or systems driven by data collected from sensors in real-time. These sophisticated computer models mirror almost every facet of a product, process, or service \cite{Tao2019Nature} with applications in Unmanned Autonomous systems \cite{Guivarch2019}, power distribution, and smart grid \cite{XieSG,Claessen2012,Angel2019}, or smart transportation \cite{Zhu2016}. The Digital Twin is supported by a multidomain simulation model built based on the knowledge about each subsystem that conforms to the real system as its constitutive physic laws, and operation experience. Considering that the Digital Twin simulation model should replicate the system's actual behavior, a systematic approach to determine the unknown parameters in the DT is required. One way is performing rigorous experimental measurements at each element of the real system, which is difficult in real-life applications. Therefore, an optimization procedure is required to find these parameters based on the available data of the system to be represented by the Digital Twin denominated behavioral matching. This procedure performs an optimization search of the best parameters for the Digital Twin subsystems to match with the real system inputs and outputs. Although the optimization problem solution can return different parameters for different operating conditions of the system, producing a family of models. 
In the literature, there are some reported results of using behavioral matching procedures combined with metaheuristic optimization algorithms like genetic algorithm, particle swarm optimization, or ant colony to find the optimal values of unknown parameters like uncertainties or controller gains with restrictions for power systems, ultra-precision machinery, optimal trajectory searching, or task scheduling \cite{Guerra2019,Bansal2019,Fang2019,Peng2019}. However, in this application, there is no criterion to choose the best possible model between a set family of models. 
\par
This paper introduces an information-based model discrimination method for Digital Twin behavioral matching results. 
The method takes the behavioral matching results obtained for different operating points of the system and calculates a set of discrete transfer function models of the Digital Twin with different complexity, which is evaluated using the information gain criterion proposed on \cite{informationGainCriterion}, the normalized Akaike information criterion (nAIC) \cite{Akaike1974}, Bayesian Information Criterion (BIC) \cite{ljung1999system}, and the minimum description length (MDL) to determine the model architecture with the best trade-off between complexity and overfitting. So, the $\nu$-gap metric \cite{zhou1998essentials} can be employed to select the best set of parameters based on the determined models of the system. The methodology is applied to assess the Digital Twin behavioral matching results performed for a real-time vision feedback infrared temperature uniformity control.
\par
The main contribution of this paper is employing information-based metrics to determine the best model during the behavioral matching process in Digital Twin applications, resulting from the presence of parametric uncertainty at different operation points of the real system.
\par
The manuscript is structured as follows. Section II introduces the information criteria. Section III presents a framework  to build a Digital Twin application. Section IV presents the study case with the Digital Twin framework application and the behavioral matching assessment using information criteria. Finally, conclusions and future works are presented.

\section{Information based metrics for behavioral matching models discrimination}
\subsection{Information gain}
Information gain proposed by \cite{informationGainCriterion} is based on the Kolmogorov complexity $K$. It is defined by \eqref{kolmogorov} for a finite sequence of letters $x$, drawn using a finite alphabet $A$, being $A$ the output alphabet of a computer $F$, with $p$ as a finite sequence of letters drawn using the input alphabet $B$ for $F$, with $l(p)$ as the length of $p$. So that the Kolmogorov Complexity is the length of the shortest program required to compute $x$.
\begin{equation}
K_{f}(x)=  \left\{
\begin{array}{ll}
\min_p  & l(p), s.t F(p)=x \\
\infty  & if ~ no ~ such ~ p ~ exist \\
\end{array} 
\right.
\label{kolmogorov}
\end{equation}

Thus, $K_f(x)$ is a suitable measure of the smallest amount of information to obtain $x$. Considering that $K(x)$ is hard to compute, it can be related with the Shannon information in a random variable \eqref{shannon}, with $H(x)$ as the entropy of $x$ and $H(x|y)$ the conditional entropy of $x$ with respect to another random variable $y$. 

\begin{equation}
J(y:x)=H(x)-H(x|y)
\label{shannon}
\end{equation}

So that, if $x$ and $y$ are random sequences from an alphabet $A$, the algorithmic information of the sequence $y$ regarding sequence $x$ is given by in terms of the Kolmogorov complexity by \eqref{algorithmInfo}, where $I(y:x)$ is a measure of how much $x$ relays on $y$ for its calculation.

\begin{equation}
I(y:x)=K(x)-K(x|y)
\label{algorithmInfo}
\end{equation}

This idea can be applied for model assessment, considering that the system observations can be divided in two datasets, one explained by the model ($x$) and another one that supports and helps to explain the first dataset ($y$). Thus, the quality of the model can be judged using a program to compute $x$ from $y$ and measure its length bounded by $K(x|y)$. As these value is lower, it indicates that the model represents better the system dynamics.  
\par
Assume a system $S$ defined as a set of $N$ input output observations $S=(u,y)$ where $u=(u_1^m,u_2^m,..,u_N^m)$, $Y=(y_1^n,y_2^n,..,y_N^n)$ for some $m,n\ge0$, $N>0$. Each pair of observations $u_i,y_i$ can be coded by a small integer $r$ representing a numeration system ($r=2$ and $r=10$ for binary and decimal).
\par
Besides, a model $F$ for the system $S(u,y)$ can be defined as a computer program $p$ that calculates the system output $y$ based on its input $u$. So that, $F$ can be defined as \eqref{computerProgram} where $C_i$ is a subset $C_i=(A_i,B_i)$ with $A_i=u_i$ and $B_i=y_i$.
\begin{equation}
F(p,i,C_i)=y_i^n,~i=0,1,...,N
\label{computerProgram}
\end{equation}
From \ref{computerProgram}, the shortest model $F$ of $S$ is the one that uses the information $I( (1,C_1),(2,C_2),...,(N,C_N):y )$ more efficiently. However, calculating $I$ from \eqref{algorithmInfo} is not possible due to the unknown of  $K_F(y)$ and $K_F(y|C)$, so only known models can be compared. For any system $S$, a trivial model $t$ can be defined from the beginning, by reading the output $y$ from a look-up table. So, for any model $p$ of $S$, the information gain $I()$ is defined by \eqref{informationGain}
\begin{equation}
I(p)=l(t)-l(p)
\label{informationGain}
\end{equation}
where $l(t)$ and $l(p)$ corresponds to the lengths of the trivial and proposed models for the system. For any model, the length is given by \eqref{modelLength}, with $L_{program}()$ as the length of the computer program that describes the model, and $L_{table}()$, as the length of the lookup table.

\begin{equation}
l()=L_{program}() + L_{table}()
\label{modelLength}
\end{equation}

In the case of $t$, the look-up table corresponds to the system outputs observations. For the model $p$, the look-up table records the difference between the system output $y$ and the estimated output $\hat{y}$ given by the model $p$,  quantifying the error or missing behavior captured by the model.
\par
To calculate the length of the look-up tables $L_{table}()$ for $t$ and $p$,  these should be codified, assuming that each element in the table corresponds to a rational number that will be scaled and represented using a numeration system $r$. So, the code-length function $l()$ for each $n$ element in the table is defined by \eqref{codeLengthFunc}, were $[~]$ represents the floor operation.

\begin{equation}
L(n)=[log_r|n|]+1
\label{codeLengthFunc}
\end{equation}

In this paper, a decimal numeration ($r=10$) system is used for look-up table codification, treating each table element $n$ as a high order integer, removing decimal period and adding the corresponding sign to $n$. For example, if $n=10.34$, it is codified as $"+1034"$ returning a length of 5, or if $n=-0.45$ its codification is $"-45"$ returning a length of 3 always removing the leading zeros. Thus, the look-up table length is given by \eqref{tableLength}.
\begin{equation}
L_{table}()=\sum_{i=1}^{n}L(i).
\label{tableLength}
\end{equation}

Likewise, to calculate the program length $l_{program}$, a similar codification rule is applied, based on the number of code lines and commands required by a programming language to implement the model of the system. According to \ref{informationGain}, some rules can be set to quantify the program length. Initially, an extended alphabet of 26 characters plus ten digits (0-9) and special symbols (\#,\%,+,-,.) are considered. Each character or digit in the code increases the length of the program by 1. However, the variable's names, as well as reserved words of the programming language, only increase the program length by 1. In \cite{informationGainCriterion}, the models were implemented using ALGOL68, but in this paper, the models will be implemented using Matlab.
\par
Notice that as the information gain of the system $I(p)$ increases, it indicates that the model $p$ offers a better explanation of the system behavior. Dividing $I(p)$ by $l(t)$ return the explanation degree of the model $p$, bounded between 0-1, where a value of 1 indicates the best level of explanation the system behavior by the model.

\subsection{Normalized Akaike information Criterion}
According to  \cite{Akaike1974}, the Akaike's Information Criterion (AIC) returns a measurement of the model quality produced by simulating a situation where the model is tested in the presence of different datasets. This criterion compares the family of models information entropy via the Kullback-Leibler divergence. Thus, the most accurate model among a family of models is the one with the smallest AIC value.
This criterion penalize the complexity of the system, it means, it will increase for systems with bigger structures and number of parameters. There are different AIC criterion forms. In this paper, the normalized AIC is calculated, which is given by \eqref{AIC}, where $N$ is the number of samples, $\epsilon(t)$ is a vector of the prediction errors, $\theta_n$ is the vector of estimated parameters, $n_y$ the number of model outputs and $n_p$ the number of estimated parameters.

\begin{eqnarray}
\label{AIC}
	nAIC=N log(det(\frac{1}{N}\sum_{1}^{N}) \epsilon(t,\hat{\theta})(\epsilon(t,\hat{\theta}))^T)+\frac{2n_p}{N}
\end{eqnarray}
On the other hand, the Bayesian Information Criterion (BIC) \cite{ljung1999system} can be calculated from AIC, which is given by\eqref{bicCriterion}
\begin{eqnarray}
		BIC=N log(det(\frac{1}{N}\sum_{1}^{N}) \epsilon(t,\hat{\theta})(\epsilon(t,\hat{\theta}))^T)\\
		\notag +N*(n_y*\log(2\pi)+1)+n_p*\log(N)
	\label{bicCriterion}
\end{eqnarray}

\subsection{Minimum Description length}
The minimum description length (MDL) is a information theory based index for evaluating model complexity, penalizing the number of parameter required to represent the system behavior \cite{pintelon2004system}. MDL can be calculated using \eqref{mdlIndex}, where $V_{ml}$ is the loss function of the model for the estimated model parameters $\hat{\theta}$, $d$ is the number of parameters in the system, and $N$ the length of the output observations vector.
\begin{equation}
	MDL=V_{ml}(\hat{\theta(z)},z)(1+\frac{d}{N})\ln(N)
	\label{mdlIndex}
\end{equation}

\subsection{$\nu$gap metric}
The Vinnicombe $\nu$gap metric \cite{zhou1998essentials} is a measurement of distance between two LTI dynamic systems P1 and P2, with right coprime factorization $P1=N_1M_1^{-1}$ and $P2=N_2M_2^{-1}$ given by \eqref{gapMetric}. It can be used as a stability indicator for robust control design. The $\nu$gap metric is always bounded between 0 and 1. As the value is close to zero the P1 and P2 are more similar with a stability margin degradation less than the $\nu$gap metric value.
\begin{eqnarray}
\scriptstyle
\delta_v(P_1,P_2)=\max_w ||(I+P_2P^*_2)^{-\frac{1}{2}}(P_1-P_2)(I+P_1P_1^*)^{-\frac{1}{2}}||_{\infty}
\label{gapMetric}
\end{eqnarray}
\par
In this paper, the $\nu$gap metric is employed to measure the similarity between the family of models resulting from the behavioral matching. for this reason, the $\nu$gap metric is calculated between each model from the behavioral matching, creating a triangular $\nu$gap matrix, which, the smaller column cumulative summation will indicate the best model.

\section{Development framework for Digital Twin applications}
\begin{figure}[h]
	\centering
	\includegraphics[width=0.4\textwidth, height=0.1\textheight]{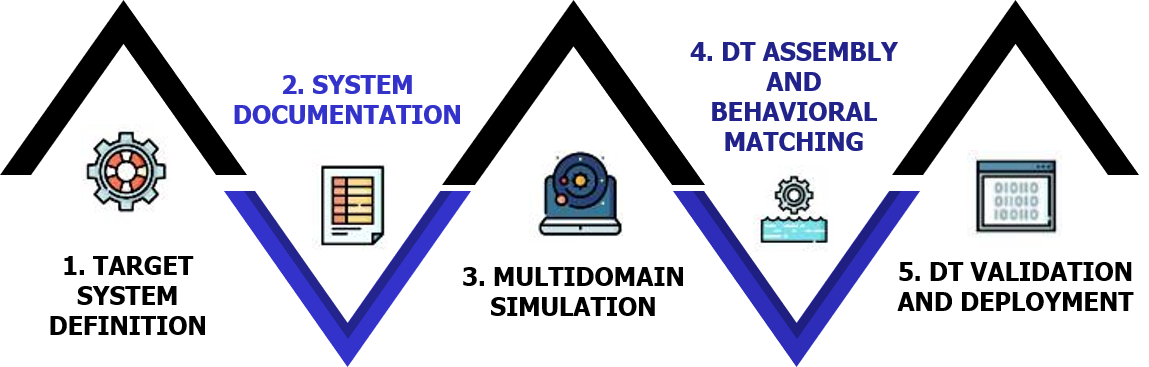}
	\caption{Digital Twin framework}
	\label{DT Framework}
\end{figure}

A framework to developing Digital Twin models is shown in Fig.\ref{DT Framework}. It is composed of five steps corresponding to the target system definition, system documentation, Multidomain simulation, DT assembly and behavioral matching, and the DT evaluation and deployment. 

In the first step, the current status of the physical system to be replicated via Digital Twin is recognized between two possible scenarios. The first one is Conceptual design, where a physical prototype is not available, and DT is employed for the initial designing task. In the second one, the physical system is operating, and DT is a supporting tool to improve system operation. In the second step,  all the available information of the system is collected to create the most accurate representation, including the control algorithms employed, Sensors and actuators datasheets, Troubleshooting and problem records, Cumulative experience of the system engineers and operators, and the system data streams. In the third step, a set of simulation models is employed to represent the real system behavior, defining the simulation domains related to the system according to the system's physical and constitutive laws as well as the appropriated computational tools for multiphysics simulation. 
\par
Once the simulation models are completed, the four-step or behavioral matching is performed. It is described in Fig.\ref{DT Behavioral Matching}, and consist of determining the unknown parameters of the system using real data collected for the system for different operating point through optimization fitting techniques like nonlinear least squares. Thus the simulation behavior of the Digital Twin is made as closely as possible to the real asset. Finally, after performing the behavioral matching, the Digital Twin is ready for the last step of real-life validation and deployment, running in parallel with the real system and being feed with live data streams to perform further analysis like prognosis or fault detection.
 \begin{figure}[h]
	\centering
	\includegraphics[width=0.45\textwidth, height=0.1\textheight]{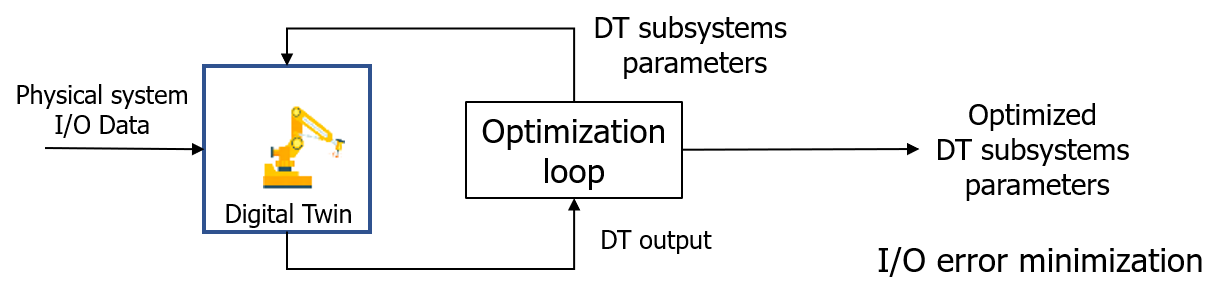}
	\caption{DT Behavorial Matching}
	\label{DT Behavoroial Matching}
\end{figure}

\section{Study case: real-time vision feedback infrared temperature uniformity control}
\begin{figure}[h]
	\centering
	\includegraphics[width=0.35\textwidth, height=0.25\textheight]{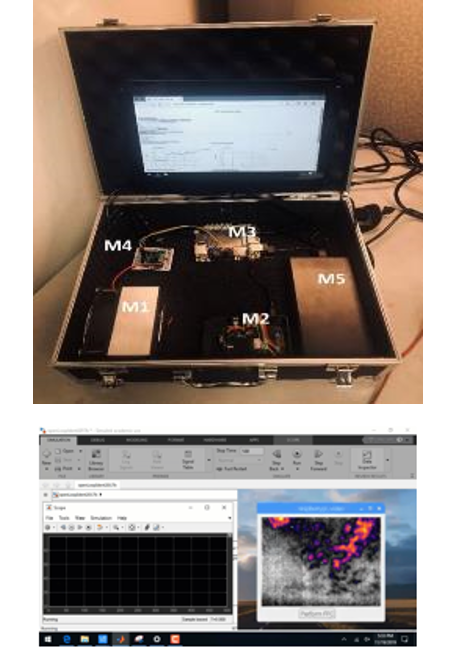}
	\caption{DT Study case: real-time vision feedback infrared temperature uniformity control}
	\label{DT Study case}
\end{figure}

The real-time vision feedback infrared temperature uniformity control presented in Fig.\ref{DT Study case} is employed in this paper as a study case for developing its Digital Twin based on the proposed framework in Section III. 
\subsection{System description, documentation and multidomain simulation developing}
As shown in Fig.\ref{DT Study case}, The system is composed of a Peltier cell (M1) that works as a heating or cooling element, a thermal infrared camera (M2) acting as a temperature feedback sensor running on a Raspberry Pi and communicated using TCP/IP communication protocol, that allows performing temperature distribution measurement and control. An additional component of the system is the LattePanda board (M3), which runs Windows 10 64-bits and executes Matlab in hardware in the loop configuration. The power applied to the Peltier cell is managed with an Arduino board (M4) via Pulse Width Modulation (PWM). The platform is equipped with its own battery (M5) that provides the power for all the system components. This study case system can be fit into the second scenario, which is open and closed-loop stable, employing a PID controller with antwindup for temperature regulation. Table.\ref{DTDocumentation} presents a summary of the properties of the system components required for the steps of multidomain simulation and behavioral matching. More details about the system implementation and real tests performed on the system can be found in \cite{Viola2019,Viola2019a}.
\begin{table}[]
	\centering
	\caption{Brief thermal system documentation}
	\label{DTDocumentation}
	\begin{tabular}{@{}cc@{}}
		\toprule
		Component                                                                            & Features                                                                                                                                                 \\ \midrule
		\begin{tabular}[c]{@{}c@{}}FLIR lepton Thread\\ Infrared thermal Camera\end{tabular} & \begin{tabular}[c]{@{}c@{}}Wavelength: 8 to 14 $\mu$m\\ Resolution: 80x60 pixels\\ Accuracy: $\pm$ 0.5$^oC$\\\\\end{tabular} \\
		\begin{tabular}[c]{@{}c@{}}TEC1-12706 \\ Peltier Module\end{tabular}                 & \begin{tabular}[c]{@{}c@{}}$Q_{max}=50W$\\ $\Delta_{Tmax}=75^oC$\\ $I_{Max}=6.4A$\\ $V_{max}=16.4V$\\\\\end{tabular}         \\
		\begin{tabular}[c]{@{}c@{}}MC33926 DC\\ Power Driver\end{tabular}                    & \begin{tabular}[c]{@{}c@{}}Input: 0-5 V\\ Output: 0-12V\\ Peak Current: 5A\\\\\end{tabular}                                  \\
		Lattepanda board                                                                     & \begin{tabular}[c]{@{}c@{}}5 inch Windows 10 64 bits PC\\ Intel Atom $\mu p$\\ 4GB of RAM\\ Built-in Arduino Leonardo board\end{tabular}                 \\ \bottomrule
	\end{tabular}
\end{table}

\begin{figure}[h]
	\centering
	\includegraphics[width=0.45\textwidth, height=0.15\textheight]{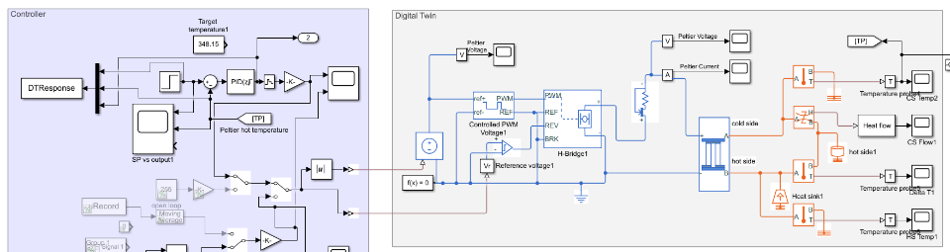}
	\caption{Assembled DT multidomain simulation}
	\label{DTTotal}
\end{figure}

The multiphysics simulation model is presented in Fig.\ref{DTTotal}. It is divided in four simulation domains. The first domain is the Electrical, composed by the power driver, the Battery and the semiconductor joint on the Peltier module. The second one corresponds to the Thermal domain defined by the heat transfer produced between the Peltier hot and cold sides, the system surface and the surroundings, and the thermal properties of the heat sink. The third domain corresponds to the fluids, given by the airflow pumped into the heat sink to keep its temperature constant. Finally, the fourth domain corresponds to the Digital Domain, composed by the PID control algorithm and the analog to digital interfaces to communicate the the control side with the thermal system. Also, this simulation domain includes the behavior of the infrared thermal camera. In this paper, the Electric, Thermal, and Digital domains will be replicated in the Digital Twin application using Matlab Simulink and Simscape as multidomain simulation packages. 
The principal component for the physical asset is the Peltier thermoelectric module which can be modeled using \eqref{peltierEqn1}-\eqref{peltierEqn3}, where $\alpha$ is the Seebeck coefficient, $R$ is the electrical resistance, $K$ is thermal conductance, $T_A,T_B$ are hot/cold side temperatures, $Q_A,Q_B$ are the hot/cold side thermal flow, and $I,V$ the applied voltage and current. Likewise, the dynamic change of the heat flow $Q$ in the hot side of the Peltier is given by \eqref{peltierEqn4}, where $C$ is the specific heat of the Peltier device and $m$ is the specific mass of the module. 
\begin{eqnarray}
Q_A=\alpha T_AI-\frac{1}{2}I^2R+K(T_A-T_B) \label{peltierEqn1}\\
Q_B=\alpha T_BI-\frac{1}{2}I^2R+K(T_B-T_A) \label{peltierEqn2} \\
V=\alpha(T_B-T_A)+IR  \label{peltierEqn3}\\
Q=Cm\frac{dT}{dt} \label{peltierEqn4}
\end{eqnarray}
\subsection{Behavioral matching}
Due to the nonlinear behavior of the Peltier module, as well as the challenge for measuring heat flow and other thermal parameters, the behavioral matching is required to determine the values of $\alpha$, $R$, $K$, and $C$. Based on the Peltier datasheet, some literature reported experimental measurements \cite{Peltier, Kubov2016}, and previous experience manipulating the system; there is possible to know the initial guess for the behavioral matching process, which are presented in Table \ref{DTPeltierParameters}.
\begin{table}[h]
	\centering
	\caption{Peltier Thermal parameters}
	\label{DTPeltierParameters}
	\begin{tabular}{@{}cccc@{}}
		\toprule
		Parameter & Datasheet    & Measurement \cite{Kubov2016} & Experience    \\ \midrule
		$\alpha$  & 53 mv        & 40 mv       & 75 mv         \\
		$R$       & 1.8 $\Omega$ & 6 $\Omega$  & 3.3 $\Omega$ \\
		$K$       & 0.5555 K/W   & 0.3333 K/W  & 0.3808 K/W    \\
		C         & 15 J/K       & 15 J/K      & 31.4173 J/K   \\ \bottomrule
	\end{tabular}
\end{table}
\par
\begin{figure}
	\centering
	\includegraphics[width=0.4\textwidth, height=0.2\textheight]{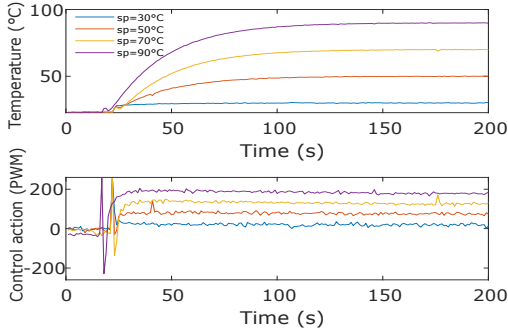}
	\caption{Peltier system responses for different steps}
	\label{modelFamily}
\end{figure}

A set of real tests is performed to acquire real data from the system, consisting of applying different step reference signals, as shown in Fig.\ref{modelFamily} to the system in order to evaluate its dynamic behavior for four different setpoints $30^oC,50^oC,70^oC$ and $90^oC$. The control signal $u$, the system output temperature $y$ and the reference signal $r$ are registered for each setpoint to determine  $\alpha$, $R$, $K$, and $C$.
\par
\begin{figure}
	\centering
	\includegraphics[width=0.4\textwidth, height=0.2\textheight]{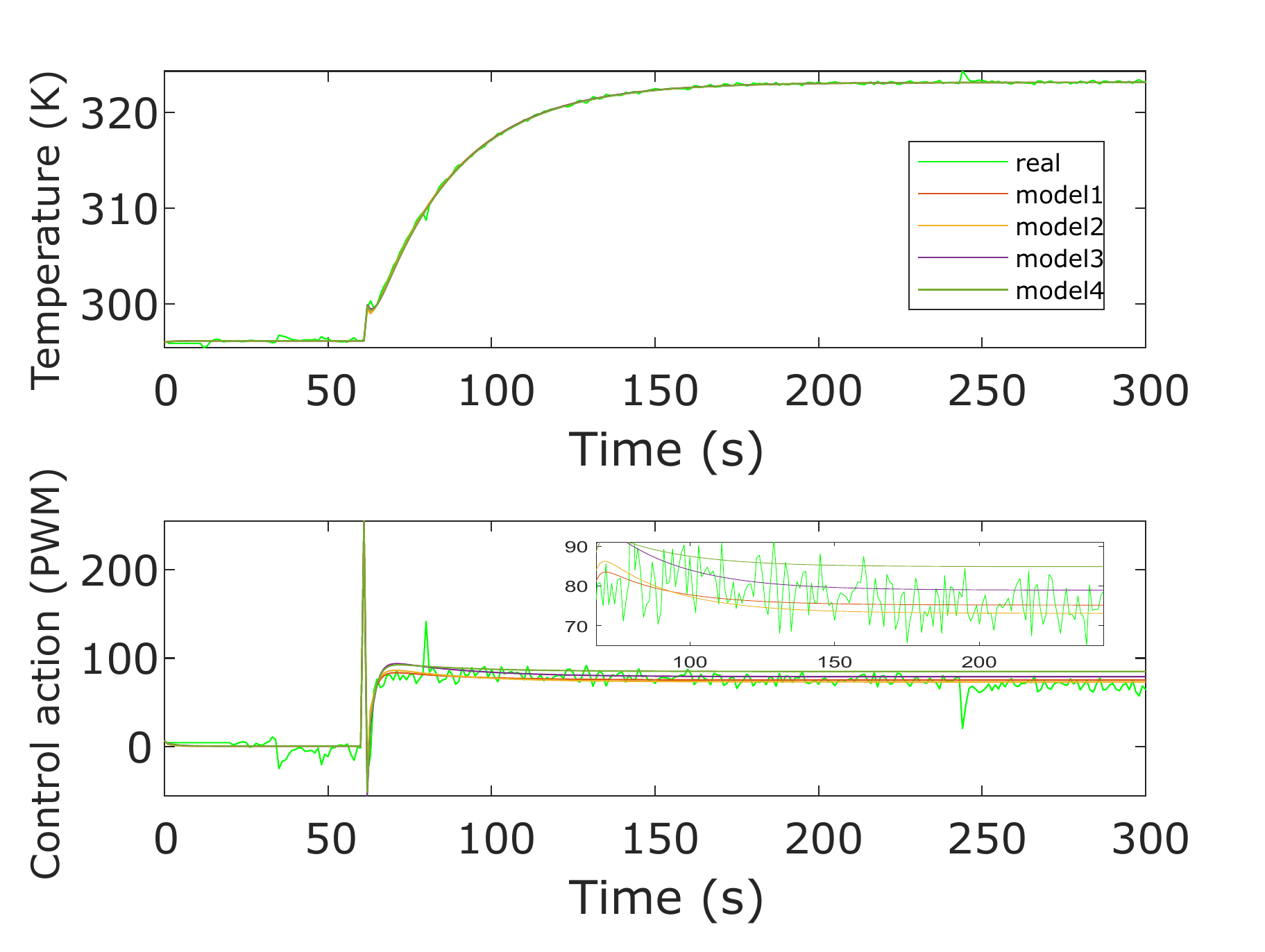}
	\caption{Digital Twin uncertainty for a setpoint of $50^oC$}
	\label{DTUncertainty}
\end{figure}

The nonlinear recursive least squares algorithm combined with the Matlab design optimization toolbox \cite{MathworksInc2020} of  is employed at each case to find the values of $\alpha$, $R$, $K$, and $C$ through matching the output and control action curves of the physical system with the Digital Twin. The sum of squared error is employed as a cost function for the parameter fitting problem defined by \eqref{LSNLCostFunction}, where $e(k)$ are the system residuals and $N$ the number of data samples. It is important to notice that $R=3.3\Omega$, which was physically measured. The obtained parameters  $\alpha$, $K$, and $C$ for each setpoint are presented in Table \ref{BMResults}. It can be observed that the Peltier thermal parameters vary among the setpoints, indicating parametric uncertainty on the system as well as a significant difference with the parameters reported in Table \ref{DTPeltierParameters}.  For example, Fig.\ref{DTUncertainty} shows the Digital Twin response for $50^oC$ setpoint with the parameters set obtained from behavioral matching registered in Table \ref{BMResults}, confirming the presence of uncertainty also in the Digital Twin. For this reason, applying model discrimination techniques is required in order to determine the most optimal and accurate set of parameters for the system Digital Twin.
\begin{equation}
F(x)=\sum_{k=0}^{N}e(k) \times  e(k)
\label{LSNLCostFunction}
\end{equation}

\begin{table}[]
	\centering
	\caption{Behavioral matching results for different setpoints}
	\label{BMResults}
	\begin{tabular}{@{}ccccc@{}}
		\toprule
		& \multicolumn{4}{c}{\textbf{Setpoint}}                                                 \\ \midrule
		\textbf{Parameter} & $30^oC$ & $50^oC$ & $70^oC$ & $90^oC$ \\
		\textbf{$\alpha$}  & $96.3mv$          & $82.5mv$           & $21.1mv$       & $29.5mv$     \\
		\textbf{$R$}       & $3.3\Omega$         & $3.3\Omega$       &$3.3\Omega$          & $3.3\Omega$          \\
		\textbf{$K$}       & $0.3 K/w$        & $0.35 K/w$     & $0.28 6K/w$       & $0.38K/w$    \\
		\textbf{$C$}       & $34.9 J/K$        & $31.93 J/K$       & $11.1 J/K$        & $13.7 J/K$              \\ \bottomrule
	\end{tabular}
\end{table}

\subsection{Digital Twin Model discrimination}
\begin{figure}[h]
	\centering
	\includegraphics[width=0.2\textwidth, height=0.1\textheight]{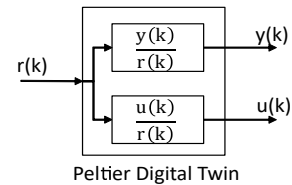}
	\caption{SIMO model for DT}
	\label{SIMOModel}
\end{figure}

The information-based metrics presented in section II are employed to perform the model discrimination assessment for the Digital Twin, which requires a model of the Digital Twin to determine the nominal set of parameters of the system. Considering that during the behavioral matching, the temperature $y$ and control $u$ action of the system was employed to determine the missing coefficients for a specific reference signal $r$, a single-input multiple-output (SIMO) system for the Digital Twin is proposed in Fig.\ref{SIMOModel}. As can be observed, it is composed by two transfer functions one between $y(k)/r(k)$ and other for $u(k)/r(k)$. The goal of this SIMO model is to consider $y$ and $u$ in the model assessment regarding the same reference signal.
\par
On the other hand, the order of the SIMO model should be in the lowest order possible in order to satisfy the Occam's razor condition, it means reducing the model complexity to avoid overfitting. For this reason, four Box-Jenkins models (BJ) given by \eqref{BJModel} are identified for $y(k)/r(k)$ and $u(k)/r(k)$ with second to fifth order polynomials for $B(z),C(z),F(z),D(z)$ for each set of parameters in Table \ref{BMResults}, conforming a 2x1 transfer function matrix. As example, Table \ref{BJ parameters} shows the  polynomial coefficients for the BJ models obtained for $y(k)/r(k)$ and $u(k)/r(k)$ using the second set of parameters for a setpoint of $50^oC$.
\begin{equation}
	y(z)=\frac{B(z)}{F(z)}u(z)+\frac{C(z)}{D(z)}e(z)
	\label{BJModel}
\end{equation}

\begin{table*}[]
	\centering
	\caption{Box-Jenkins models family for the behavioral matching results at $50^oC$}
	\label{BJ parameters}
	\begin{tabular}{@{}cccccc@{}}
		\toprule
		\multicolumn{2}{c}{\textbf{Polynominal}}          & \textbf{Order 22221} & \textbf{Order 33331}  & \textbf{Order 44441}          & \textbf{Order 55551}               \\ \midrule
		\multirow{4}{*}{$\frac{y(k)}{r(k)}$} & \textbf{B} & 0 0.03 -0.028        & 0 -0.001 0.002 0      & 0 -0.011 0.01 0 0             & 0 -0.045 -0.603 1.245 -0.601 -0.01 \\
		& \textbf{C} & 1 -0.817 0.002       & 1 1.235 0.609 -0.016  & 1 -0.048 -0.007 -0.695 -0.001 & 1 2.098 1.155 -0.075 -0.085 0.047  \\
		& \textbf{D} & 1 -1.772 0.786       & 1 0.247 -0.613 -0.634 & 1 -1.035 0.041 -0.689 0.683   & 1 1.094 -0.966 -1.31 -0.033 0.214  \\
		& \textbf{F} & 1 -1.997 0.999       & 1 -2.949 2.899 -0.95  & 1 -1.806 -0.158 1.735 -0.77   & 1 -1.776 1.612 -1.067 0.392 -0.125 \\
		&            & \textbf{}            & \textbf{}             & \textbf{}                     & \textbf{}                          \\
		\multirow{4}{*}{$\frac{u(k)}{r(k)}$} & \textbf{B} & 0 0 0                & 0 0 0 0               & 0 -0.006 0.011 -0.006 0.002   & 0 0.089 -0.217 0.144 0.013 -0.029  \\
		& \textbf{C} & 1 0.003 0            & 1 0.127 -0.077 -0.001 & 1 0.142 -0.372 -0.297 -0.02   & 1 -0.161 0.017 0.607 -0.233 -0.057 \\
		& \textbf{D} & 1 -0.995 -0.006      & 1 -0.87 -0.204 0.074  & 1 -0.84 -0.563 0.121 0.282    & 1 -1.156 0.971 -1.34 0.716 -0.186  \\
		& \textbf{F} & 1 -1.978 0.978       & 1 -2.219 1.455 -0.237 & 1 -2.264 1.117 0.557 -0.41    & 1 -2.365 2.259 -2.21 2.148 -0.832  \\ \bottomrule
	\end{tabular}
\end{table*}

Now, the model discrimination criteria are calculated for the identified SIMO system for each set of parameters presented in Table \ref{BMResults}. In the case of Information Gain, each BJ model is evaluated as a difference equation employing only the transfer function part of \eqref{BJModel}. From \eqref{informationGain}, the information gain is given by the difference between the trivial $l(t)$ model and the BJ model $l(BJ)$. Likewise, the length of each program is calculated as the sum of the lengths of the computer program plus the look-up table \eqref{modelLength}. In the case of the trivial program, its length $l(t)$ is calculated using the coding rules proposed in section 2, which is implemented in Matlab with a length of 15, being the same for all the trivial models. Regarding the look-up table for the trivial model $t$, it is coded using the rules in section 2, and its length depends on each real setpoint response.
\par
The implementation of BJ models is also performed in Matlab with a length of $l(BJ)=176$. Considering that the same code works for any of the proposed BJ models, the code length $l(BJ)$ keeps constant at each calculation. Regarding the look-up table, it is calculated as $y-\hat{y}$, where $y$ is the physical system response, and $\hat{y}$ is the response obtained from each BJ model evaluated. Again, its length depends on $y-\hat{y}$ and is calculated using the rules in section 2.
\par
Finally, the total Information Gain of the SIMO model is calculated as the sum of the individual information gains from $y(k)/r(k)$ and $u(k)/r(k)$. In this case, the most suitable model is the one with the higher information gain, it means, the one that provides more information about the system. The trivial and BJ models codes can be found in \textit{https://github.com/tartanus/Information-Gain-Criterion}.
\par
Considering that only one criterion may not be enough to choose the most suitable model for the system, the nAIC, BIC, and MDL information gain criteria are calculated for the SIMO system, using the expressions (9)-(12). Table \ref{DTInformationDiscrim} shows the calculation of the information criteria for each MISO BJ model regarding its corresponding dataset. As can be observed, the Information gain shows that for setpoints $50^oC$ and  $90^oC$, a second-order BJ model is enough to represent the system dynamics, while for setpoints  $30^oC$ and  $70^oC$, models of third and fourth-order are more representative for that specific datasets. It is important to say that the Information Gain method is sensitive to the decimal precision of the measurements as well as the look-up table.

On the other hand, it can be noticed that using the nAIC, BIC, and MDL criteria, the second model BJ order is the best model to represent the system dynamics. So, we can say that based on the multiple assessment metrics employed, a second-order BJ model represents the Digital Twin dynamic with the best trade-off between complexity and overfitting.
\par
Once the best type of SIMO model for the Digital Twin is selected, the next step consists of determining the nominal set of parameters of the Digital Twin, that works for multiple operating points. In that sense, the $\nu$Gap metric is calculated for the second-order BJ models obtained for each operating point. Thus, the set of parameters with the less cumulative $\nu$Gap metric determines the nominal set of parameters, considering that $\nu$gap metric measures the distance between the models based on the $H_\infty$ norm seeking presented in \eqref{gapMetric}. The obtained result of the $\nu$gap metric for the second-order BJ models are 2.93, 2.74, 2.22, and 2.28 for the $30^oC$,$50^oC$,$70^oC$, and $90^oC$ setpoints respectively. It can be observed that the smallest value of $\nu$gap metric is given for the third set of parameters corresponding to a setpoint of $70^oC$. So that, we can say that these values of $\alpha,R,K,C$ correspond to the nominal operation parameters for the Digital Twin.

\begin{table*}[]
		\label{DTInformationDiscrim}
		\setlength\abovecaptionskip{0\baselineskip}
		\setlength{\belowcaptionskip}{-4pt}
		\caption{Information criterion calculation for Digital Twin model assessment}
		\centering
		\small\addtolength{\tabcolsep}{-4pt}
	\begin{tabular}{cccccccccccccccccc}
		\hline
		\multirow{2}{*}{\textbf{SP}} & \multirow{2}{*}{\textbf{\begin{tabular}[c]{@{}c@{}}model\\ order\end{tabular}}} & \multicolumn{3}{c}{\textbf{y(k)/r(k)}}          & \multicolumn{3}{c}{\textbf{u(k)/r(k)}}          & \multirow{2}{*}{\textbf{\begin{tabular}[c]{@{}c@{}}IGT\\ (u,y)\end{tabular}}} & \multirow{2}{*}{\textbf{\begin{tabular}[c]{@{}c@{}}nAIC \\ y(k)/r(k)\end{tabular}}} & \multirow{2}{*}{\textbf{\begin{tabular}[c]{@{}c@{}}nAIC \\ u(k)/r(k)\end{tabular}}} & \multirow{2}{*}{\textbf{nAICT}} & \multirow{2}{*}{\textbf{\begin{tabular}[c]{@{}c@{}}BIC \\ y(k)/r(k)\end{tabular}}} & \multirow{2}{*}{\textbf{\begin{tabular}[c]{@{}c@{}}BIC \\ u(k)/r(k)\end{tabular}}} & \multirow{2}{*}{\textbf{BIC}} & \multirow{2}{*}{\textbf{\begin{tabular}[c]{@{}c@{}}mdl\\  y(k)/r(k)\end{tabular}}} & \multirow{2}{*}{\textbf{\begin{tabular}[c]{@{}c@{}}mdl \\ u(k)/r(k)\end{tabular}}} & \multirow{2}{*}{\textbf{\begin{tabular}[c]{@{}c@{}}mdl \\ Total\end{tabular}}} \\
		&                                                                                 & \textbf{l(t)} & \textbf{l(BJ)} & \textbf{IG(y)} & \textbf{l(t)} & \textbf{l(BJ)} & \textbf{IG(u)} &                                                                               &                                                                                     &                                                                                     &                                 &                                                                                    &                                                                                    &                               &                                                                                    &                                                                                    &                                                                                \\ \hline
		\multirow{4}{*}{30}          & 22221                                                                           & 1242          & 681            & 561            & 1019          & 1014           & 5              & 566                                                                           & 310.24                                                                              & 330.39                                                                              & \textbf{640.63}                 & 310.24                                                                             & 330.39                                                                             & \textbf{640.63}               & 0.27                                                                               & 0.27                                                                               & \textbf{0.54}                                                                  \\
		& 33331                                                                           & 1242          & 741            & 501            & 1019          & 1055           & -36            & 465                                                                           & 472.21                                                                              & 383.82                                                                              & 856.03                          & 472.21                                                                             & 383.82                                                                             & 856.03                        & 0.34                                                                               & 0.34                                                                               & 0.68                                                                           \\
		& 44441                                                                           & 1242          & 637            & 605            & 1019          & 998            & 21             & \textbf{626}                                                                  & 819.77                                                                              & 464.58                                                                              & 1284.35                         & 819.77                                                                             & 464.58                                                                             & 1284.35                       & 0.49                                                                               & 0.49                                                                               & 0.97                                                                           \\
		& 55551                                                                           & 1242          & 1219           & 23             & 1019          & 1078           & -59            & -36                                                                           & 429.28                                                                              & 467.93                                                                              & 897.21                          & 429.28                                                                             & 467.93                                                                             & 897.21                        & 0.48                                                                               & 0.48                                                                               & 0.96                                                                           \\ \hline
		\multirow{4}{*}{50}          & 22221                                                                           & 2048          & 979            & 1069           & 1622          & 1493           & 129            & \textbf{1198}                                                                 & 0.78                                                                                & 0.80                                                                                & \textbf{1.58}                   & 1283.65                                                                            & 1290.03                                                                            & \textbf{2573.68}              & 2.52                                                                               & 2.52                                                                               & \textbf{5.03}                                                                  \\
		& 33331                                                                           & 2048          & 1082           & 966            & 1622          & 1506           & 116            & 1082                                                                          & 0.84                                                                                & 0.84                                                                                & 1.68                            & 1327.87                                                                            & 1325.32                                                                            & 2653.19                       & 2.75                                                                               & 2.75                                                                               & 5.50                                                                           \\
		& 44441                                                                           & 2048          & 1067           & 981            & 1622          & 1549           & 73             & 1054                                                                          & 0.86                                                                                & 0.84                                                                                & 1.70                            & 1358.30                                                                            & 1349.62                                                                            & 2707.92                       & 2.88                                                                               & 2.88                                                                               & 5.77                                                                           \\
		& 55551                                                                           & 2048          & 1836           & 212            & 1622          & 1653           & -31            & 181                                                                           & 0.93                                                                                & 0.93                                                                                & 1.87                            & 1404.68                                                                            & 1404.46                                                                            & 2809.15                       & 3.33                                                                               & 3.33                                                                               & 6.67                                                                           \\ \hline
		\multirow{4}{*}{70}          & 22221                                                                           & 2898          & 1286           & 1612           & 2890          & 2042           & 848            & 2460                                                                          & 1.82                                                                                & 2.24                                                                                & \textbf{4.06}                   & \textbf{2337.72}                                                                   & \textbf{2543.94}                                                                   & \textbf{4881.66}              & 10.40                                                                              & 10.40                                                                              & \textbf{20.80}                                                                 \\
		& 33331                                                                           & 2898          & 1258           & 1640           & 2890          & 1712           & 1178           & \textbf{2818}                                                                 & 1.91                                                                                & 2.26                                                                                & 4.17                            & 2406.90                                                                            & 2579.63                                                                            & 4986.53                       & 11.13                                                                              & 11.13                                                                              & 22.27                                                                          \\
		& 44441                                                                           & 2898          & 1429           & 1469           & 2890          & 2107           & 783            & 2252                                                                          & 1.94                                                                                & 2.17                                                                                & 4.11                            & 2446.20                                                                            & 2561.95                                                                            & 5008.16                       & 10.61                                                                              & 10.61                                                                              & 21.21                                                                          \\
		& 55551                                                                           & 2898          & 2591           & 307            & 2890          & 2013           & 877            & 1184                                                                          & 1.97                                                                                & 2.29                                                                                & 4.26                            & 2485.83                                                                            & 2642.01                                                                            & 5127.84                       & 15.18                                                                              & 15.18                                                                              & 30.37                                                                          \\ \hline
		\multirow{4}{*}{90}          & 22221                                                                           & 2721          & 1219           & 1502           & 2712          & 1930           & 782            & \textbf{2284}                                                                 & 1.88                                                                                & 2.30                                                                                & \textbf{4.18}                            & 2226.30                                                                            & 2417.66                                                                            & \textbf{4643.96}              & 11.06                                                                              & \textbf{11.06}                                                                     & \textbf{22.11}                                                                 \\
		& 33331                                                                           & 2721          & 1519           & 1202           & 2712          & 1904           & 808            & 2010                                                                          & 2.28                                                                                & 2.29                                                                                & 4.57                            & 2434.50                                                                            & 2437.36                                                                            & 4871.86                       & 11.43                                                                              & 11.43                                                                              & 22.86                                                                          \\
		& 44441                                                                           & 2721          & 1386           & 1335           & 2712          & 2014           & 698            & 2033                                                                          & 1.90                                                                                & 2.28                                                                                & 4.18                 & 2284.10                                                                            & 2457.17                                                                            & 4741.27                       & 11.78                                                                              & 11.78                                                                              & 23.57                                                                          \\
		& 55551                                                                           & 2721          & 1170           & 1551           & 2712          & 2143           & 569            & 2120                                                                          & 4.17                                                                                & 1.91                                                                                & 6.08                            & 3354.17                                                                            & 2312.17                                                                            & 5666.34                       & 12.38                                                                              & 12.38                                                                              & 24.77                                                                          \\ \hline
	\end{tabular}
\end{table*}

\section{Conclusions and future works}
In this paper, a model discrimination methodology was introduced for Digital Twin assessment based on information criteria indices and $\nu$gap metric. The procedure is employed to determine the most suitable parameters during the behavioral matching process of Digital Twin in the presence of parametric uncertainty for different operation points. A SIMO transfer function model is employed to represent the overall behavior of the Digital Twin, choosing the most suitable model based on multiple information indices to define a model with the best trade-off between complexity and overfitting. Thus, the $\nu$gap metric can be applied to determine the best set of parameters based on the optimal models of the Digital Twin. The model assessment performed for the Digital Twin for a real-time vision feedback infrared temperature uniformity control system shows that the estimated parameters are closer to the values reported by the manufacturer. However, its correct estimation is required to obtain a correct Digital Twin representation of the physical system. As future works, the introduction of different statistical methods like maximum likelihood, fisher information, and stochastic assessment techniques is proposed to improve the results of this method and make it more general for its application into much more complex systems.

\bibliographystyle{ieeetr}
\bibliography{reference1}

\end{document}